\documentclass[aps,prl,twocolumn,floatfix,english,showpacs,10pt,superscriptaddress,longbibliography]{revtex4-2}

\usepackage{graphicx}
\usepackage{booktabs}
\usepackage{CJKutf8}
\usepackage{amsmath}
\usepackage{physics}
\usepackage{amssymb}
\usepackage{mathrsfs}
\usepackage{colordvi}
\usepackage{verbatim}
\usepackage{xcolor}
\usepackage{multirow}
\usepackage{mathrsfs}
\usepackage{epsfig}
\usepackage{lipsum}
\usepackage{enumitem}
\usepackage{amsfonts}

\usepackage[unicode=true, breaklinks=false, pdfborder={0 0 1}, backref=false,
colorlinks=true, linkcolor=blue, urlcolor=blue, citecolor=blue]{hyperref}%
\usepackage{cleveref}
\providecommand{\U}[1]{\protect\rule{.1in}{.1in}}

\begin{document}
\begin{CJK}{UTF8}{gbsn}
\title{Finite-size effects on the edge loss probability in non-Hermitian quantum walks}
\author{Shuaixian Liu}


\affiliation{Hunan Key Laboratory for Super-Microstructure and Ultrafast Process, School of Physics, Central South University, 932 South Lushan Road, Changsha 410000, People’s Republic of China}

\author{Yulan Dong}
\affiliation{School of Microelectronics and Physics, Hunan University of Technology and Business, Changsha 410205, China}

\author{Bowen Zeng}
\email{\textcolor{black}{zengbowen@csust.edu.cn}}
\affiliation{Hunan Provincial Key Laboratory of Flexible Electronic Materials Genome Engineering,
School of Physics and Electronic Sciences, Changsha University of Science and Technology, Changsha 410114, People’s Republic of China}

\author{Meng-Qiu Long}
\email{\textcolor{black}{mqlong@csu.edu.cn}}
\affiliation{Hunan Key Laboratory for Super-Microstructure and Ultrafast Process, School of Physics, Central South University, 932 South Lushan Road, Changsha 410000, People’s Republic of China}

\makeatletter
\newcommand{\rmnum}[1]{\romannumeral #1}
\newcommand{\Rmnum}[1]{\expandafter\@slowromancap\romannumeral #1@}
\makeatother
\begin{abstract}
A dynamical bulk–edge relation in quantum walks has been theoretically proposed and experimentally observed, in which a power-law dependence of the bulk loss probability is associated with a pronounced peak of loss probability at the edge. This behavior has been proven to arise from imaginary gap closing and the non-Hermitian skin effect in the infinite limit without boundary effects. However, in a finite-size chain, we find that boundary scattering can suppress this edge burst. Meanwhile, imaginary gap opening, together with the
non-Hermitian skin effect, can also induce a large loss probability at the edge. Our results provide insights into
finite-size quantum dynamics.

\end{abstract}
\maketitle

\section{Introduction}
Bulk-boundary correspondence, a milestone in condensed matter physics, refers to nontrivial bulk topological invariants associated with the system's eigenfunction corresponding to the emergence of edge states~\cite{su1980soliton,berry1984quantal,heeger1988solitons,zak1989berry,hasan2010colloquium,qi2011topological,bernevig2013topological,asboth2016short}. In the non-Hermitian crystals, where the non-Hermitian terms arise from their interaction with the environment, the nonreciprocal coupling reshapes such bulk-boundary correspondence by highlighting the role of the boundary~\cite{hatano1996localization,lee2016anomalous,xiong2018,kunst2018biorthogonal,yao2018edge,zhou2026critical}. For example, nonreciprocal coupling leads to the  accumulation of bulk states at the boundary, called the non-Hermitian skin effect (NHSE)~\cite{longhi2019probing,kawabata2019symmetry,okuma2020,ashida2020non,zhang2022review,lin2023topological,wang2024amoeba,zeng2023radiation,wu2025hybrid,hu2025acoustic}, which alters the integral range of the allowed wave vector from the conventional Brillouin zone (BZ) to the generalized Brillouin zone (GBZ) for the calculation of bulk topological invariants~\cite{yokomizo2019non,yao2018edge,yang2020non}. Intriguingly, the non-Hermitian skin effect itself corresponds to nontrivial bulk spectral winding number, another kind of bulk-boundary correspondence associated with the system's eigenvalue~\cite{okuma2020,ding2022non,zhang2020correspondence,bergholtz2021exceptional,okuma2023non,yu2024non,yang2025inverse}.  

A recent dynamical bulk-edge scaling relation for the loss probability when quantum walks in a lossy lattice are considered~\cite{xue2022non} can also be considered a kind of bulk-boundary correspondence in a certain sense. With an excitation, such a scaling relation refers to the fact that the power-law dependence of the bulk loss probability on the distance from the observation point to the excitation point results in a prominent peak for the edge loss probability, called the non-Hermitian edge burst~\cite{wang2021quantum,xue2022non,hu2023steady,ren2023dissipation,xiao2024observation,zhu2024observation,wen2024investigation,yuce2024strong,sen2025edge}.
Based on the Green's function, Xue \textit{et al.}~\cite{xue2022non} demonstrated that the non-Hermitian edge burst results from the interplay between the imaginary gap closing of the spectrum under periodic boundary conditions (PBCs) and the NHSE. Here, the PBC spectrum is employed solely for the purpose of analyzing the imaginary gap closing or opening. The imaginary gap closing, which is determined solely by the Hermitian part of the Hamiltonian~\cite{ma2024imaginary}, is associated with a power-law dependence of the bulk loss probability. The NHSE in this model is realized by dissipation and flux~\cite{PhysRevLett.125.186802,li2022gain,wu2022flux}, which can be related to nonreciprocal coupling via a unitary transformation~\cite{yang2025inverse}.  The NHSE governs the directionality of the quantum walks and assists in trapping the wave packet upon its arrival at the boundary, as verified by time-dependent perturbation theory~\cite{wen2024investigation}. Experimentally, the non-Hermitian edge burst has been observed from the photonic quantum walk~\cite{xiao2024observation,zhu2024observation} to classical-wave metamaterials~\cite{zou2026non}.

The understanding of the edge burst from the above studies is based solely on bulk properties, where boundary scattering is negligible, making this analysis less comprehensive.  In practice, however, the boundary plays a crucial role in non-Hermitian dynamics. Under an excitation and before the wave packet reaches the boundary, the response associated with the time-domain Green's function or wave packet evolution is independent of the boundary condition~\cite{mao2021boundary,zhou2024abnormal,xue2025}. However, once the excitation point is placed at the boundary, the wave packet dynamics acquire an extra factor due to boundary effects~\cite{yang2025real}. Moreover, depending on the strength of the boundary confinement, the wave packet may either be trapped at the edge or reflected back into the bulk~\cite{xue2025}. Therefore, the manifestation
of the non-Hermitian edge burst, as a dynamical bulk-edge
scaling relation associated with non-Hermitian dynamics, in
finite-size systems must be strongly influenced by boundary
effects.

In this work, we focus on the impact of the boundary effects on the edge loss probability in a finite-size lossy lattice by tuning the non-Hermitian on-site dissipation. At moderate dissipation, a strong NHSE accompanied by weak boundary scattering enables efficient trapping of the wave packet at the boundary. 
This boundary-induced trapping, together with the NHSE and imaginary gap closing, gives rise to the non-Hermitian edge burst. In contrast, in the regime of extremely strong dissipation, analyses based on the Lyapunov exponent imply that boundary scattering becomes dominant, thereby suppressing the edge burst. Intriguingly, even when the imaginary gap of the PBC spectrum remains open, extreme dissipation drives bulk modes to those under the imaginary gap closing regime. Combined with the NHSE, this can still lead to a significant loss probability at the boundary.

\section{Edge loss probability in the finite-size chain}

\begin{figure}
    \centering    \includegraphics[width=0.45\textwidth]{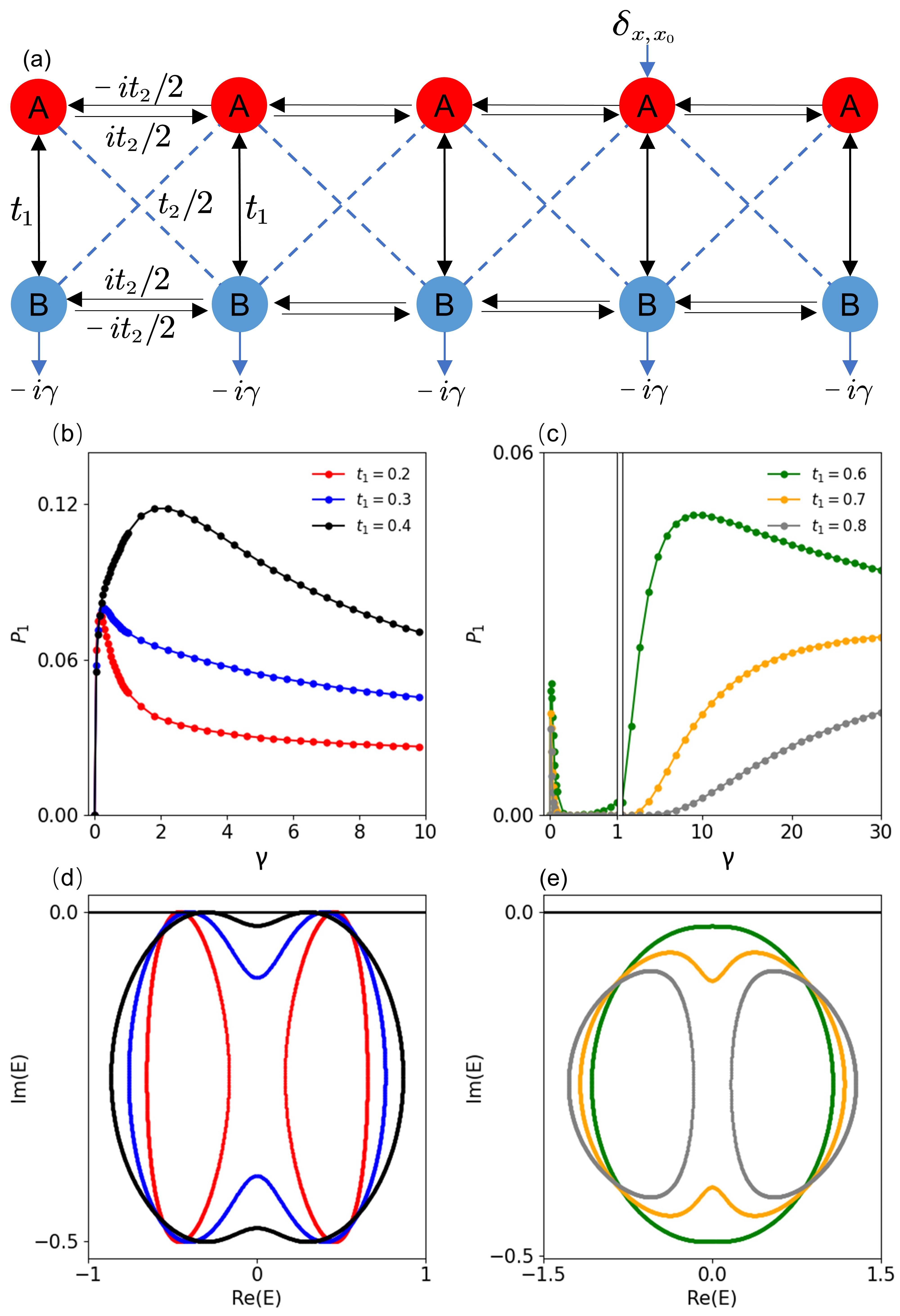} 
    \caption{(a) Two coupled chains, with $t_1$ and $t_2$ being the coupling parameters and $i\gamma$ representing the on-site dissipation on chain \textit{B}. Under excitation $\delta_{x,x_{0}}$, taking a value of 1 at $x=x_0$ and 0 elsewhere, the variation of the edge loss probability with $\gamma$ for (b) $t_1 < t_2$ and (c)  $t_1 > t_2$,   with the parameters $t_2 = 0.5$, size $L=100$, and excitation location $x_0=90$. The corresponding PBC spectrum for (d) $t_1 < t_2$ and (e) $t_1 > t_2$ and $\gamma=0.5$, where the gap closing and opening correspond to whether the PBC spectrum touches the real axis or not. }
    \label{fig-1}
\end{figure}

We consider the same ladder model as Refs~\cite{wang2021quantum,xue2022non}, as shown in Fig.~\ref{fig-1}(a); it is composed of two chains, \textit{A} and \textit{B}. The
loss rate of $ i\gamma$ is introduced exclusively on chain \textit{B}, which serves as the only non-Hermitian term in this system. The quantum dynamics in such a lossy lattice is governed by  
\begin{gather}\label{eq-dynamic}
i\frac{d\psi_x^A}{dt} = t_1\psi_x^B + i\frac{t_2}{2}(\psi_{x-1}^A - \psi_{x+1}^A) + \frac{t_2}{2}(\psi_{x-1}^B + \psi_{x+1}^B), \nonumber\\
i\frac{d\psi_x^B}{dt} = t_1\psi_x^A - i\frac{t_2}{2}(\psi_{x-1}^B - \psi_{x+1}^B) + \frac{t_2}{2}(\psi_{x-1}^A + \psi_{x+1}^A) \nonumber\\ -i\gamma\psi_x^B,\nonumber \\
\end{gather}
with $\psi$ being the amplitude. The Hamiltonian in momentum space reads 
\begin{equation}\label{eq-Hmomen}
    H(\beta)  =  \left(
    \begin{array}{cc}
       \frac{i t_2}{2} \beta^{-1} -  \frac{i t_2}{2} \beta &  \frac{ t_2}{2} \beta^{-1} + t_1 + \frac{ t_2}{2}\beta\\
       \frac{ t_2}{2} \beta^{-1} + t_1 + \frac{ t_2}{2}\beta & \frac{-i t_2}{2} \beta^{-1} +  \frac{i t_2}{2} \beta -i \gamma \\
    \end{array}
  \right),
\end{equation}
with diagonal and nondiagonal terms representing the interchain and intrachain couplings. Here, $\beta=e^{ik}$ denotes the wave vector. For a $\delta$-type excitation $\psi_x^A = \delta_{x,x_0}$, the particles can escape from chain \textit{B} during the quantum walks. The particle loss rate can be estimated as $(d/dt) \langle \psi | \psi \rangle = i \langle \psi | ( H^{\dagger} - H ) | \psi \rangle = -\sum_{x} 2\gamma \left| \psi_{x}^{B} \right|^{2}$~\cite{wang2021quantum}, where $H$ is the Hamiltonian in real space. Therefore, the total loss probability at location $x$ reads 
\begin{equation}
P_x=2\gamma\int_0^\infty |\psi_x^B(t)|^2 dt.\label{eq-loss}
\end{equation}
The corresponding loss probability at the edge is denoted by $P_1$. Within an appropriate parameter range, edge burst, {i.e.}, $P_1 \gg P_{\text{min}}$ can occur, where $P_{\text{min}}= \text{min}\{P_1,P_2,\cdots,P_{x_0}\}$ is the minimum of $P$ between the edge and excitation point. Numerically, for a single-site excitation, we sum the loss probability over all time steps until the deviation of the total accumulated loss from unity is below 0.1\%, at which point the integration time  for the above equation is considered sufficient.

Xue \textit{et al.}~\cite{xue2022non} demonstrated that the appearance of an edge burst in such a model requires the combination of the non-Hermitian skin effect and imaginary gap closing. Here, the strength of the NHSE, characterized by the radius of the GBZ $r_{G}$, can be obtained by solving the eigenequation of Eq.~\eqref{eq-Hmomen},
\begin{equation}
\label{eq-eigenvalue}
    E^2 + i \gamma E - \frac{\gamma t_2}{2}\left( \beta - \beta^{-1}\right) -t_2^2 -t_1^2 - t_1 t_2 \left(\beta + \beta^{-1}\right)=0.
\end{equation}
Here, $E$ denotes the eigenenergy. For any $E$, there are two roots $\beta_1$ and $\beta_2$ that satisfy $\beta_1\beta_2 = (t_1-\gamma/2)/(t_1+\gamma/2)$. According to the GBZ theory under open boundary conditions (OBCs)~\cite{yokomizo2019non,yao2018edge,yang2020non}, these roots must satisfy $|\beta_1|=|\beta_2|$, which gives $r_{G}=\sqrt{|(t_1-\gamma/2)/(t_1+\gamma/2)|}$. 
While the expression for the GBZ radius seems independent of $t_2$, that does not mean $t_2$ is irrelevant. In fact, if $t_2=0$, the energy $E$ in the above equation becomes independent of $\beta$, in which case the NHSE cannot occur. From another perspective, $t_2$ is introduced as an effective flux, together with the on-site dissipation, and $t_1$ allows the PBC spectrum to form a closed loop in the complex energy plane, as shown in Fig.~\ref{fig-1}(d) and ~\ref{fig-1}(e), giving rise to a nontrivial spectral winding number for the OBC spectrum. This winding number underlies the
existence of the NHSE ~\cite{okuma2020,zhang2020correspondence}. Therefore, for finite $t_1$ and $t_2$, any nonzero dissipation $\gamma$ causes the GBZ to deviate from the BZ, implying the appearance of the NHSE.

The imaginary gap closing refers to a scenario where, among the eigenmodes corresponding to the Bloch wave vector $\beta=e^{ik}$ with $k\in[0,2\pi]$, specific modes whose eigenvalues are purely real with no imaginary component exist. If the imaginary gap is closed, an eigenenergy $\omega_0$ corresponding to $k_0$ with zero imaginary part must exist. Substituting real $\omega_0$ into Eq.~(4), we have 
\begin{equation}
    \omega_0^2+i \gamma \omega_0-i \gamma t_2  \sin k_0-t_2^2-t_1^2-2t_1 t_2 \cos k_0=0,
\end{equation}
where the imaginary part  $- i \gamma \omega_0 +  i \gamma  t_2 \sin k_0 = 0$ and real part $\omega_0^2  -t_2^2-t_1^2-2t_1 t_2 \cos k_0=0$. From the imaginary part we obtain $\omega_0 = t_2 \sin k_0$. Substituting this into the real part leads to $(t_1 + t_2 \cos k_0)^2 = 0$.
For the Bloch wave vector,  the existence of a solution requires $|t_1|\leqslant|t_2|$, that is, the condition for imaginary gap closing. From a physical intuition perspective, whenever the quantum walks traverses lossy chain \textit{B}, dissipation inevitably occurs. Therefore, the condition for imaginary gap closing requires that the nondissipation modes  walk on only chain \textit{A} and the effective intrachain coupling between \textit{A} and \textit{B} vanishes~\cite{ma2024imaginary}, namely, $t_1 + t_2 \cos k =0$, or, equivalently, $ |t_1| \leqslant |t_2| $. In Fig.~\ref{fig-1}(d) and ~\ref{fig-1}(e), we plot the spectrum under periodic boundary conditions for different $t_1$ and fixed $t_2$. For $|t_1|\leqslant |t_2|$, modes exist that touch the real axis without any imaginary component  [see Fig.~\ref{fig-1}(d)]. However, for $|t_1| > |t_2|$, the spectrum always maintains a finite imaginary gap, as shown in Fig.~\ref{fig-1}(e).

It should be noted that the above two conditions for edge burst are discussed for an infinite-size chain, where boundary effects are negligible. 
In a finite-size chain, however, the loss probability at the edge exhibits a much more intricate behavior, as shown in  Fig.~\ref{fig-1}(b) and ~\ref{fig-1}(c), where the numerical dependence of $P_1$ on $\gamma$ for different $t_1$ with length $L=100$ is plotted. Here, the edge burst in finite-size systems can still be defined as $P_1\gg P_{\text{min}}$.  For $|t_2|\geqslant |t_1|$, $P_1$ vanishes at $\gamma=0$ due to the absence of the NHSE. 
As $\gamma$ increases, the curves of $P_1$ for different $t_1$ initially grow rapidly with nearly identical slopes, reach a maximum, and then gradually decrease. In particular, $P_1$ becomes small for large $\gamma$. Moreover, a larger $t_1$ leads to a higher peak value of $P_1$. Even for $|t_2|<|t_1|$, corresponding to the condition of imaginary gap opening, a sizable loss probability at the edge can still emerge, as shown in Fig.~\ref{fig-1}(c). In this case, each curve  contains two segments of growth and decay, giving rise to two distinct peaks. The first segment occurs in a narrow interval of $\gamma$, while the second rise appears after the decay and reaches a higher peak as $\gamma$ increases further.

Before proceeding with a detailed analysis, we highlight two key distinctions between the finite chain and the infinite chain. First, the coexistence of the NHSE and imaginary gap closing does not necessarily guarantee that $P_1$ remains much larger than $P_{\text{min}}$. Second, within an appropriate range of $\gamma$, the NHSE together with imaginary gap opening can also induce the edge burst.

\section{Edge loss probability under the condition of imaginary gap closing}\label{se:IGC}

To estimate the boundary effects in a finite-size chain, we begin by reviewing the theory for edge burst developed by Xue \textit{et al.}~\cite{xue2022non}. The amplitude $\psi_x^B(t)$
corresponds to the projection coefficients of the time-evolution operator acting on the initial excitation on chain \textit{A} and can be written as $\psi_x^B(t) = \langle x,B | \Theta(t)e^{-i H t} | x_0, A \rangle$. It is often expressed in terms of the Green’s function as $\psi_x^B(t) = i \langle x,B | G(t) | x_0, A \rangle$, with $G(t)= - i \Theta(t)e^{-i H t}$. In the frequency domain, the Green’s function takes the form $G(t) = 1/(2\pi) \int_{-\infty}^{\infty} G(\omega) e^{-i\omega t}d\omega  $, with $G(\omega) = 1/(\omega -H + i0^{+})$. It should be noted that $G(t)$ does not depend on the boundary condition
before the wave packet reaches the boundary, although the frequency-domain Green’s function $G(\omega)$ does
depend on the boundary condition. Consequently, while using the Bloch basis under PBCs or non-Bloch
basis under OBCs modifies the value of $G(\omega)$, it does not affect the resulting $G(t)$ for bulk dynamics.
This important property has been demonstrated in several theoretical works~\cite{mao2021boundary,zhou2024abnormal,xue2025}. Substituting the above relations in Eq.~\eqref{eq-loss} then yields
\begin{equation}\label{eq-Gw}
    P_x = \frac{\gamma}{\pi}\int_{-\infty}^\infty d\omega|\langle x,B|G(\omega)|x_0,A\rangle|^2.
\end{equation}
For this two-band model, the matrix element can be estimated by inserting the Bloch basis, yielding 
\begin{equation}
    \langle x,B|G(\omega)|x_0,A\rangle=\oint_{|\beta|=1}\frac{d\beta \beta^{x-x_0}}{2\pi i\beta}\frac{t_1+t_2(\beta+\beta^{-1})/2}{\text{det}[\omega+i0^+-H(\beta)]}.
\end{equation}
According to Eq.~\eqref{eq-eigenvalue}, there are two roots, denoted by $\beta_L(\omega)$ and $\beta_R(\omega)$. The introduction of dissipation in Eq.~\eqref{eq-Hmomen} ensures that both roots always satisfy $|\beta_L(\omega)|\geqslant1\geqslant|\beta_R(\omega)|$ for real $\omega$~\cite{xue2022non}. With the residue theorem, Eq.~\eqref{eq-Gw} can be written as 
\begin{equation}\label{eq-P1}
  P_x=\frac{\gamma}{\pi}\int_{-\infty}^{\infty} d\omega |f_{L/R}(\omega)|^2|\beta_{L/R}(\omega)|^{2(x-x_0)}  ,
\end{equation}
with 
\begin{equation}
\begin{aligned}
f_L(\omega) &= \frac{t_1 + t_2 \frac{\beta_L + \beta_L^{-1}}{2}}{t_2(t_1 + \gamma/2)(\beta_L - \beta_R)}, \quad x \leqslant x_0, \\
f_R(\omega) &= \frac{t_1 + t_2 \frac{\beta_R + \beta_R^{-1}}{2}}{t_2(t_1 + \gamma/2)(\beta_R - \beta_L)}, \quad x \geqslant x_0.
\end{aligned}
\end{equation}
Here, the subscript $L$ and $R$ represents leftward and rightward propagation, respectively. 
It is evident that the bulk loss probability decreases with increasing relative distance between the observation point and excitation point $|x-x_0|$. Consequently, the dominant contribution to Eq.~\eqref{eq-P1} arises from the region near the minimum modulus of $\beta_{L}(\omega)$ for leftward propagation. For $x\leqslant x_0$, $\beta_{L}(\omega)$ can take the minimum modulus $|\beta_{L}(\omega_0)|=1$ for a certain real $\omega_0$ only if the nondiagonal coupling in Eq.~\eqref{eq-Hmomen} vanishes~\cite{ma2024imaginary}. Under this condition, the Hermitian part of the Hamiltonian becomes block diagonal. Mathematically, this requirement yields 
$t_1+t_2 \cos k_0 = 0$ and $t_2 \sin k_0 = \omega_0$. This condition explains why closing of the imaginary gap requires $|t_1| \leqslant |t_2|$ for the PBC spectrum. However, when this nondiagonal term vanishes, $f_{L}(\omega)$ also becomes zero. Therefore, to evaluate Eq.~\eqref{eq-P1}, it is necessary to perform an expansion around $\omega_0$ such that
$|f_L(\omega)|^2\approx Q\delta \omega^m$ and $|\beta_L(\omega)| \approx 1+K\delta \omega^n \approx e^{K\delta \omega^n}$, with $\delta \omega = \omega - \omega_0$, $n=m=2$, and  expansion coefficients 
\begin{equation}
    \begin{aligned}
    Q &=\frac{\sin^2k_0}{t_1^2(\gamma^2\cos^2k_0+4t_1^2\sin^2k_0)}, \\ 
    K& =\frac{1}{2} \mathrm{Re} [\frac{\gamma \omega_0}{t_1^3(2\omega_0+i\gamma)}]=\frac{\omega_0^2}{t_1^3}\frac{\gamma}{(4\omega_0^2+\gamma^2)}. 
    \end{aligned}
\end{equation}
Then Eq.~\eqref{eq-P1} can be approximated as 
\begin{align}\label{eq-bulk}
P_x=\frac{2\gamma}{\pi}\Gamma(\frac{3}{2})Q(2K)^{-\frac{3}{2}}|x-x_0|^{-\frac{3}{2}}.
\end{align}

\begin{figure}
    \centering    \includegraphics[width=1\linewidth]{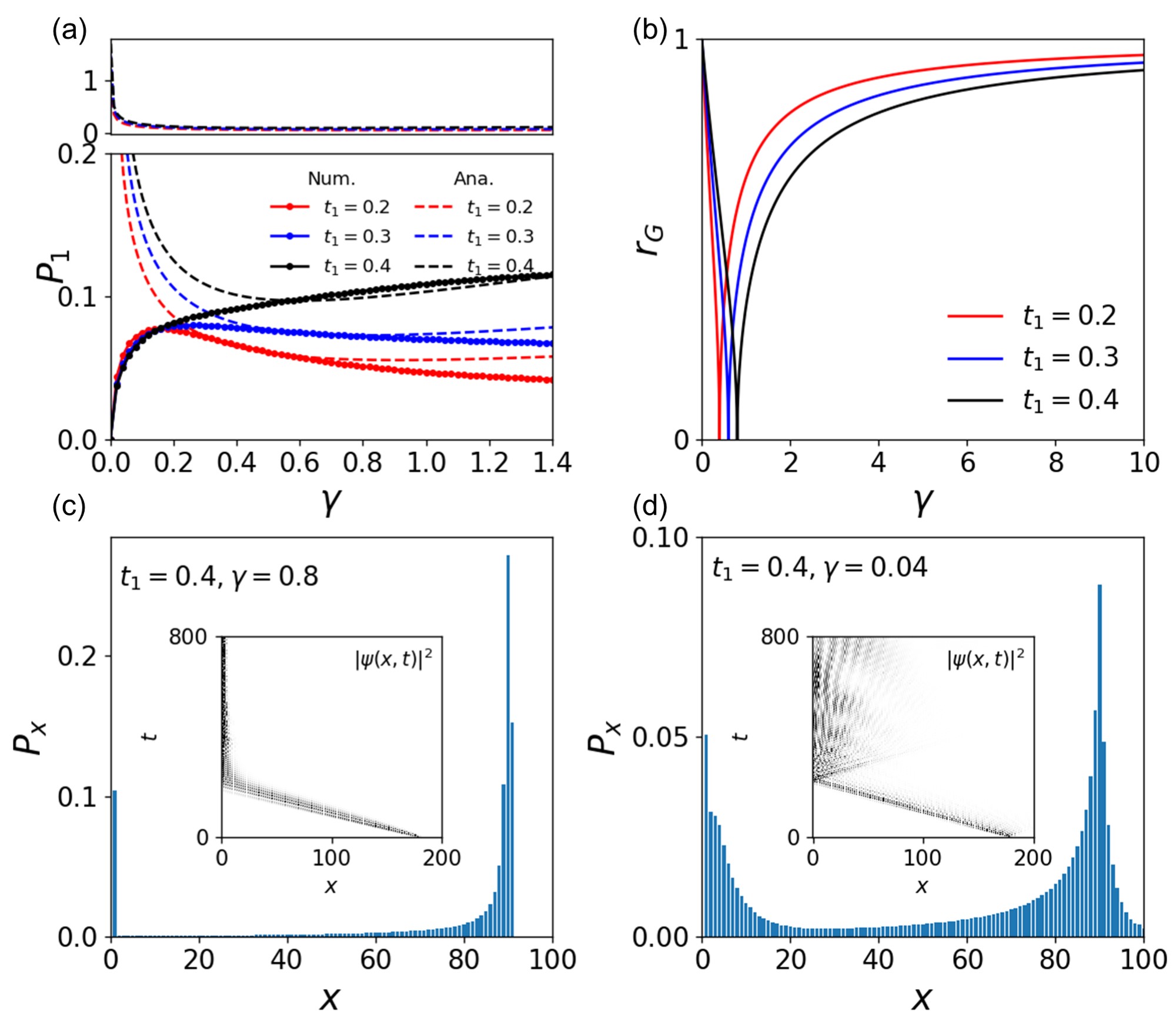}
    \caption{(a)  Comparison of the analytical results obtained from Eq.~\eqref{eq-edge} (dashed lines) and the numerical results of $P_1$ for different $t_1$. (b) Radius of GBZ $r_G$ as a function of $\gamma$. Under(c)weak scattering and (d) strong scattering, the distribution of loss probability. The evolution of the wave packet  temporally and spatially is shown in the insets. Common parameters are $t_2=0.5,L=100$ and $x_0=90$.  }
    \label{fig-2}
\end{figure}

Intuitively, one might expect the edge loss probability to be small. 
However, Eq.~\eqref{eq-bulk} also describes the dissipation for the negative values of $x$, which are completely blocked and may be trapped by the boundary, thus potentially causing an edge burst~\cite{xue2022non}. From this perspective, the edge loss probability can be estimated as 
\begin{equation}\label{eq-edge}
    \begin{aligned}
        &P_1 \approx \sum_{x=
    {-\infty}}^0P_x^\infty  \approx  \int_{-\infty}^0 dx P_x^\infty\\
    &\approx  \frac{4\gamma}{\pi}\Gamma(\frac{3}{2})Q(2K)^{-\frac{3}{2}}x_0^{-\frac{1}{2}}. \\
    \end{aligned}
\end{equation}
It is evident that  $P_1 \gg P_{\text{min}}$. 
Figure~\ref{fig-2}(a) presents a comparison between  Eq.~\eqref{eq-edge} and the numerical results for the edge loss probability [Fig.~\ref{fig-1}(b)] in a finite-size chain. They agree well in certain regions; however, for both large and small $\gamma$, they exhibit opposite trends. 

For a given $\gamma$, both the numerical and analytical results increase as $t_1$ becomes larger. This trend can be understood by writing $P_1$ as a function of $t_1$ as 
\begin{equation}
     P_1(t_1)\propto(t_2^2-t_1^2)^{-1/2} \ t_1^{1/2}\ [\gamma^2+4(t_2^2-t_1^2)]^{1/2}.
\end{equation}
Similarly, the behavior of Eq.~\eqref{eq-edge} with respect to 
$\gamma$ can be analyzed by simplifying the expression as
\begin{align}\label{eq-eapp}
    P_1(\gamma)\propto \gamma^{-1/2}(4\omega_0^2+\gamma^2)^{1/2}.
\end{align}
When $\gamma\ll2\omega_0$, this expression exhibits a power-law decay with an exponent of 
−1/2. In contrast, when $\gamma\gg2\omega_0$, $P_1$ follows a power-law increase with an exponent of 
$1/2$, as shown by the analytical curve in Fig.~\ref{fig-2}(a). The minimum of Eq.~\eqref{eq-eapp} occurs at $\gamma = 2\omega_0$,  obtained from the condition $\partial  P_1(\gamma) /\partial \gamma =0$. In addition, Eq.~\eqref{eq-edge} also leads to  unphysical results, $P_1(\gamma) > 1$, for extremely small $\gamma$ [see Fig.~\ref{fig-2}(a)] and sufficiently large $\gamma$ (not shown) since $ P_1(\gamma)$ must be less than 1 due to the inevitable bulk dissipation.  

The discrepancy between analytical results based on the infinite-size chain and numerical results can be attributed to boundary scattering. 
An intuitive way to estimate this scattering is through the strength of the NHSE, which governs the drift of the wave packet and suppresses scattering. As shown in Fig.~\ref{fig-2}(b), the GBZ radius first decreases and then increases as $\gamma$ increases. When the GBZ radius approaches 1 at either small or large 
$\gamma$, strong scattering occurs as the left-moving wave packet reaches the boundary. By contrast, a radius approaching zero corresponds to a pronounced skin effect and thus weak scattering. In particular, the GBZ radius becomes zero at $t_1 = \gamma/2$, coinciding with the non-Bloch $\mathcal{PT}$ transition~\cite{hu2024geometric,xiong2024graph}, which will be discussed later. 

Direct evidence of scattering effects is shown in Fig.~\ref{fig-2}(c) and~\ref{fig-2}(d). When the scattering is negligible, the spatial distribution $P_x$ exhibits  $P_1 > P_{\min} $ and $ P_3 > P_2$, following the power law of bulk loss probability [see Eq.~\eqref{eq-bulk}]. The leftward-moving wave packet becomes trapped at the boundary, as shown in the inset of Fig.~\ref{fig-2}(c). In contrast,  
for small 
$\gamma$, where strong edge scattering occurs, $P_1 > P_{\min}$ still holds, but $P_2 > P_3$. 
The observation $P_2>P_3$ serves as a clear signature of strong boundary scattering since it deviates from the power-law decay in Eq.~\eqref{eq-bulk}. The inset of Fig.~\ref{fig-2}(d) further shows that part of the wave packet is reflected at the boundary and subsequently propagates rightward.  

Returning to Fig.~\ref{fig-2}(a), the region where the analytical and numerical curves coincide corresponds to the weak-scattering regime. In contrast, the region where the two curves deviate corresponds to the strong-scattering regime.
 
\begin{figure}
    \centering    \includegraphics[width=1\linewidth]{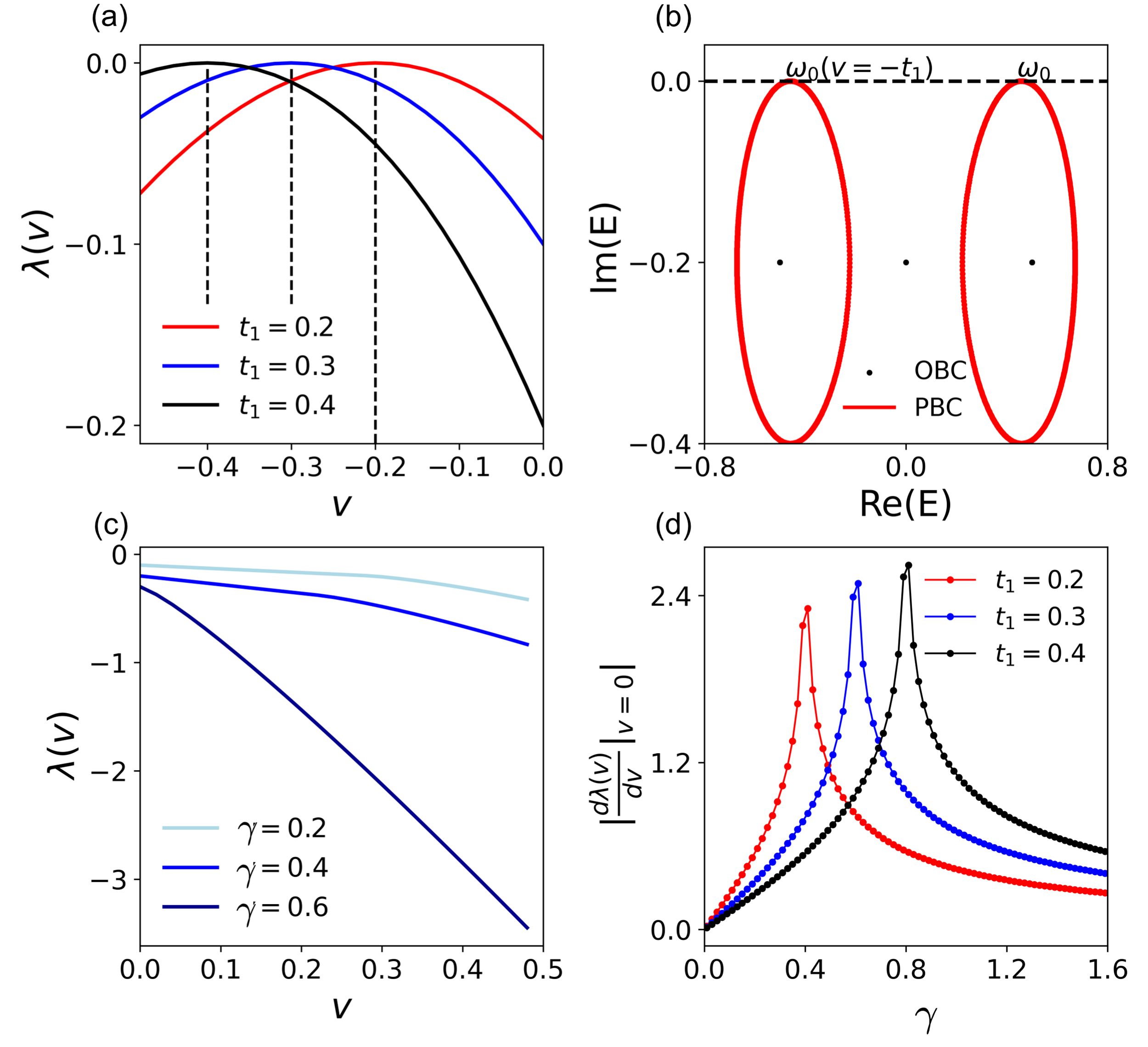}    \caption{Lyapunov exponent  for (a) negative velocity  for different $t_1$ and fixed  $\gamma=1$ before the wave packet reaches the boundary and (c) positive velocity for different $\gamma$ and fixed  $t_1=0.3$ after it arrives at the boundary. The extreme point at $t_1=0.2$ in (a) corresponds to $\omega_0$ on the PBC spectrum (red lines) in (b). (d) The variation of $\left| \frac{d  \lambda (v)}{d v }|_{v=0} \right|$ with $\gamma$ for different $t_1$. }
    \label{fig-3}
\end{figure}

The motion and reflection of the wave packet can be characterized more precisely using the Lyapunov exponent~\cite{longhi2019probing,xue2025,yang2025real}
\begin{equation}
    \lambda (v)= \text{Im}\left[E(k^s)-k^s v\right],
\end{equation}
where $v$ denotes the velocity of wave packet. $E(k)$ is obtained from Eq.~\eqref{eq-eigenvalue} by substituting $\beta=e^{ik}$, and $\omega(k^s)$ is the eigenvalue of $v$-dependent saddle points, which satisfies $\frac{d \text{Re}[E(k)]}{d k}|_{k=k^s} = v$ and $\frac{d \text{Im}[E(k)]}{d k}|_{k=k^s} = 0$. By empolying the Lefschetz–thimble method~\cite{witten2011analytic,MukherjeePhysRevB.90.035134,kanazawa2015structure,yang2025real} to identify the effective saddle points, we obtain the Lyapunov exponent associated with leftward propagation, as shown in Fig.~\ref{fig-3}(a). For a given initial excitation, the dynamics of the wave packet are dominated by the maximum of the Lyapunov exponent, and the corresponding velocity is determined by the condition $\frac{d  \lambda (v)}{d v } = 0$. 
The negative velocity indicates leftward propagation, and its magnitude increases with increasing $t_1$. This behavior is consistent with the numerical results of $P_1(t_1)$ in Fig.~\ref{fig-2}(a) with fixed $\gamma$ for different $t_1$ since a larger velocity allows the wave packet to reach the boundary more rapidly, thereby reducing dissipation in the bulk while enhancing dissipation at the boundary.

The calculation of the Lyapunov exponent $\lambda(v) $ does not depend on the boundary conditions. However, for the extremal points of $\lambda(v)$, the associated wave vectors are real and lie within the BZ~\cite{xue2025}. In particular, several real wave vectors have been reported to satisfy the condition $\frac{d \text{Im}[E(k)]}{d k}|_{k=k^s} = 0$. As shown in the PBC spectrum in Fig.~\ref{fig-3}(b), $\omega_0$ corresponds to one of the extreme point for $\text{Im}[E(k)]$. By combining the derivative of Eq.~\eqref{eq-eigenvalue} with respect to $k$ and the above discussions on $v$-dependent saddle points, we have 
\begin{equation}
    2 \omega_0 v + i \gamma v - i \gamma t_2 \cos k_0 + 2 t_1 t_2 \sin k_0 = 0.
\end{equation} 
Taking the real part of this relation yields
\begin{equation}
    \omega_0 v + t_1 t_2 \sin k_0=0,
\end{equation}
from which the velocity is found to be
$v=-t_1$. This value corresponds precisely to the extreme point of $\lambda(v)$ in Fig.~\ref{fig-3}(a). The group velocity associated with $\omega_0$ can also be used to determine the direction of the edge burst.

Having utilized Fig.~\ref{fig-3}(a) to analyze the leftward propagation of the wave packet before it reaches the boundary, we now turn to the scattering after it arrives at the boundary. It should be noted that when the wave packet reaches the boundary, only Lyapunov exponents with $v \geqslant 0 $ need to be considered because the boundary blocks leftward propagation. Consequently, even in the presence of a strong non-Hermitian skin effect (which favors leftward accumulation), the relevant saddle point on the OBC spectrum with $v=0$ governs the wave packet evolution near the boundary, As shown in Fig.~\ref{fig-3}(c). Nevertheless, the wave packet can also propagate to the right side due to a significant distribution of rightward-propagating velocities near the saddle point. By analogy with the density of states~\cite{10.1063/5.0129590}, $\frac{d  \lambda (v)}{d v }|_{v=0} $ is expected to quantify the probability of positive velocity, {i.e.}, the probability of scattering, as shown in Fig.~\ref{fig-3}(d). A large value of $\left| \frac{d  \lambda (v)}{d v }|_{v=0} \right|$ implies a large difference between $\lambda (v)$ for the rightward-propagating velocity and saddle point and thus a low scattering probability. Consequently, the scattering probability first decreases and then increases as $\gamma$ increases, which is consistent with the strength of the non-Hermitian skin effect shown in Fig.~\ref{fig-2}(b). The largest $\left| \frac{d  \lambda (v)}{d v }|_{v=0} \right|$ occurs when $r_G=0$. Under these parameters, the OBC spectrum collapses to a single point, as illustrated in Fig.~\ref{fig-3}(b). This corresponds to a non-Bloch $\mathcal{PT}$ transition for the non-Hermitian energy band; however, that is merely a coincidence, as the $\mathcal{PT}$ transition has no intrinsic connection to the scattering probability~\cite{hu2024geometric}.

\section{Edge loss probability under the condition of imaginary gap opening}

\begin{figure}
    \centering   \includegraphics[width=1\linewidth]{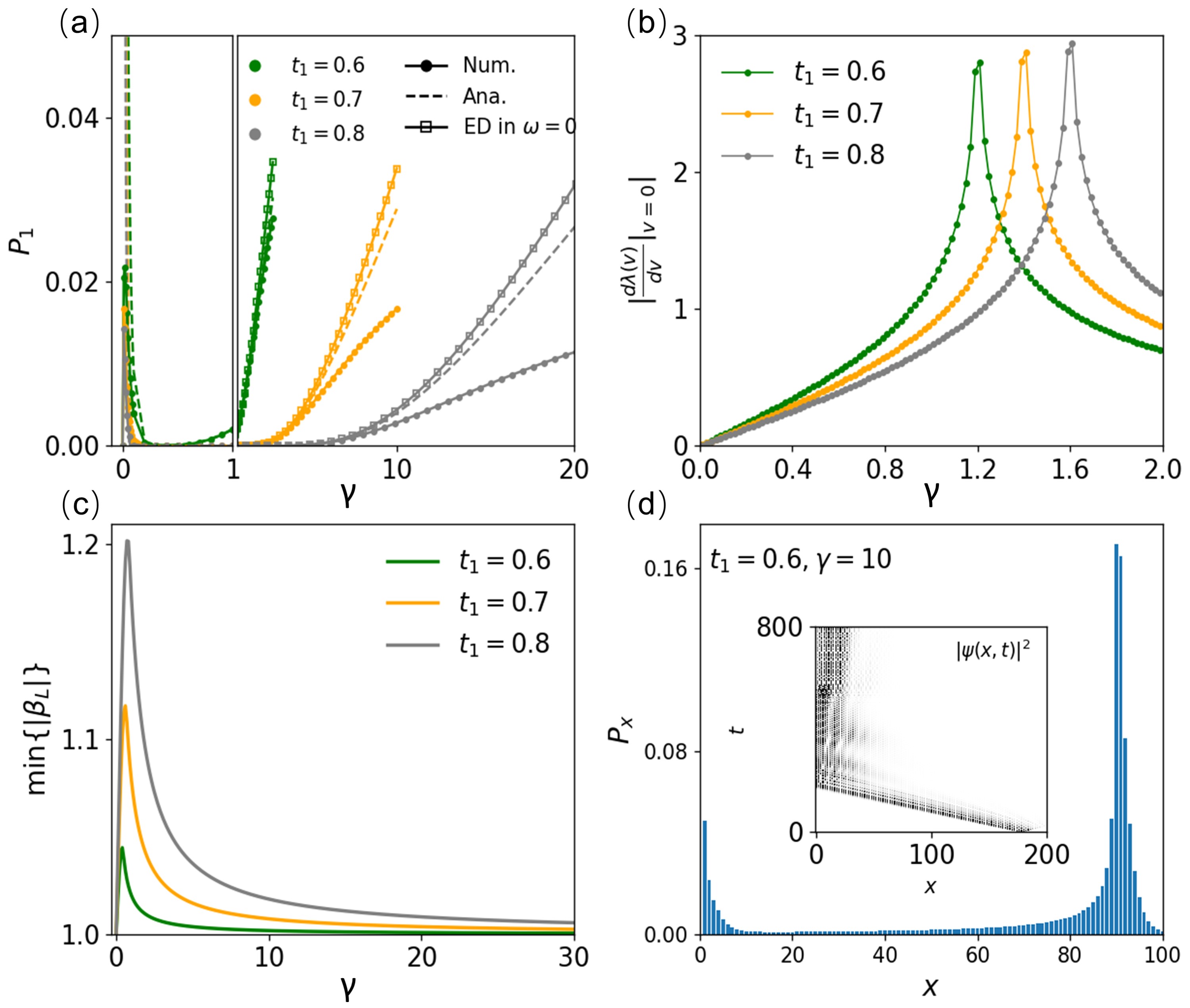}
    \caption{(a) Under the condition of imaginary gap opening, a
    comparison of the edge loss probability obtained from numerical
    results (dots) and analytic results (dashed lines) and estimated based
    on an expansion around $\omega=0$ (squares). Variation of (b) $\left| \frac{d  \lambda (v)}{d v }|_{v=0} \right|$ and (c) $\text{min}[|\beta_L(\omega)|]$ with $\gamma$. (d) The appearance of an  edge burst under a large $\gamma=10$ under strong scattering. Common parameters are  $t_2=0.5,L=100$, and $x_0=90$.}
    \label{fig-4}
\end{figure}

For the imaginary gap opening without considering the boundary scattering, the edge loss is estimated as 
\begin{equation}\label{eq-igo}
   P_1 \approx \sum_{x=
    -\infty}^{0}\frac{\gamma}{\pi}\int_{-\infty}^{\infty}d\omega |f_L(\omega)|^2|\beta_L(\omega)|^{2(x-x_0)}.
\end{equation}
The analytical and numerical results are compared in Fig.~\ref{fig-4}(a). They exhibit good agreement over an intermediate range of $\gamma$ but deviate at both small and large $\gamma$. Similar to the discussion of Fig.~\ref{fig-2}(a), this behavior can be attributed to negligible boundary scattering in the agreement regime and pronounced boundary scattering in the deviation regime. This distinction is captured by the magnitude of
 $\left| \frac{d  \lambda (v)}{d v }|_{v=0} \right|$, which is large in the former regime and small in the deviation regime, as shown in Fig.~\ref{fig-4}(b).

For $|t_1|>|t_2|$, an imaginary gap always exists between the PBC spectrum and the real axis. In other words, for any real 
$\omega$, $|\beta_L(\omega)|>1$ and $|f_L(\omega)|>0$ always hold. Consequently, the asymptotic behavior of the integral is dominated by the neighborhood of $\text{min}[|\beta_L(\omega)|]$. As shown in Fig.~\ref{fig-4}(c), as $\gamma$ increases, $\text{min}[|\beta_L(\omega)|]$ first deviates from 1 and then gradually approaches 1. A higher value of 
$\text{min}[|\beta_L(\omega)|]$ corresponds to a more rapidly decaying bulk mode.

If we compare Figs. ~\ref{fig-4}(a) and~\ref{fig-4}(c), it is evident that the agreement regime corresponds to a large value of $\text{min}[|\beta_L(\omega)|]$. Our numerical results indicate that $\omega=0$ is the minimum point of $\gamma$-dependent $|\beta_L(\omega)|$. It can be proved that $\omega=0$ is always an extreme point of $|\beta_L(\omega)|$ as a function of $\gamma$. Specifically, the derivative of $|\beta_L(\omega)|^2$ with respect to $\omega$ is given by
\begin{equation}
\frac{d|\beta_L(\omega)|^2}{d\omega}=2\text{Re}\beta_L(\omega) \frac{d\text{Re}\beta_L(\omega)}{d\omega}+2\text{Im}\beta_L(\omega)\frac{d\text{Im}\beta_L(\omega)}{d\omega}.
\end{equation}
At $\omega=0$, $\beta_L(0)= \frac{-t_1^2-t_2^2 - \sqrt{(t_1^2-t_2^2)^2+t_2^2\gamma^2}}{2t_2(t_1+\gamma/2)}$ is purely real, with $\text{Im}\beta_L(0)=0$. Additionally, from Eq.~\eqref{eq-eigenvalue}, we have 
\begin{equation}
    \frac{d\beta}{d\omega}=\frac{(2\omega+i\gamma)}{2t_1t_2+t_2\gamma+(t_1^2+t_2^2-\omega^2-i\gamma \omega)/\beta}, 
\end{equation}
which is purely imaginary at $\omega=0$, indicating $\frac{d\text{Re}\beta_L(\omega)}{d\omega}|_{\omega=0}=0$; thus, $\frac{d|\beta_L(\omega)|^2}{d\omega}|_{\omega=0}=0$. 

We now perform a second-order expansion around the neighborhood of $\text{min}[|\beta_L(0)|]$. Since $\frac{d^2\beta}{d\omega^2}|_{\omega=0}$ is real, we have 
\begin{equation}
    |\beta_L(\omega)|^2 = \left|\beta + \beta' \omega + \frac{\beta''\omega^2}{2} \right|^2\approx \beta^2 + (\beta \beta'' - \beta'^2) \omega^2. 
\end{equation}
$f_L(\omega)$, $f_L'(\omega)$ and $f_L''(\omega)$ at $\omega=0$  are real, purely imaginary,
and real. Similarly, $|f_L(\omega)|^2  \approx f^2 + (f f'' - f'^2) \omega^2$. The numerical results derived by substituting these expansions in Eq.~\eqref{eq-igo} are illustrated in fig.~\ref{fig-4}(a). It can be seen that $P_1$ under weak scattering can be estimated in the neighborhood
of $\omega=0$, analogous to the dominant $\omega_0$ under the condition of imaginary gap closing. 

\begin{figure}
    \centering   \includegraphics[width=1\linewidth]{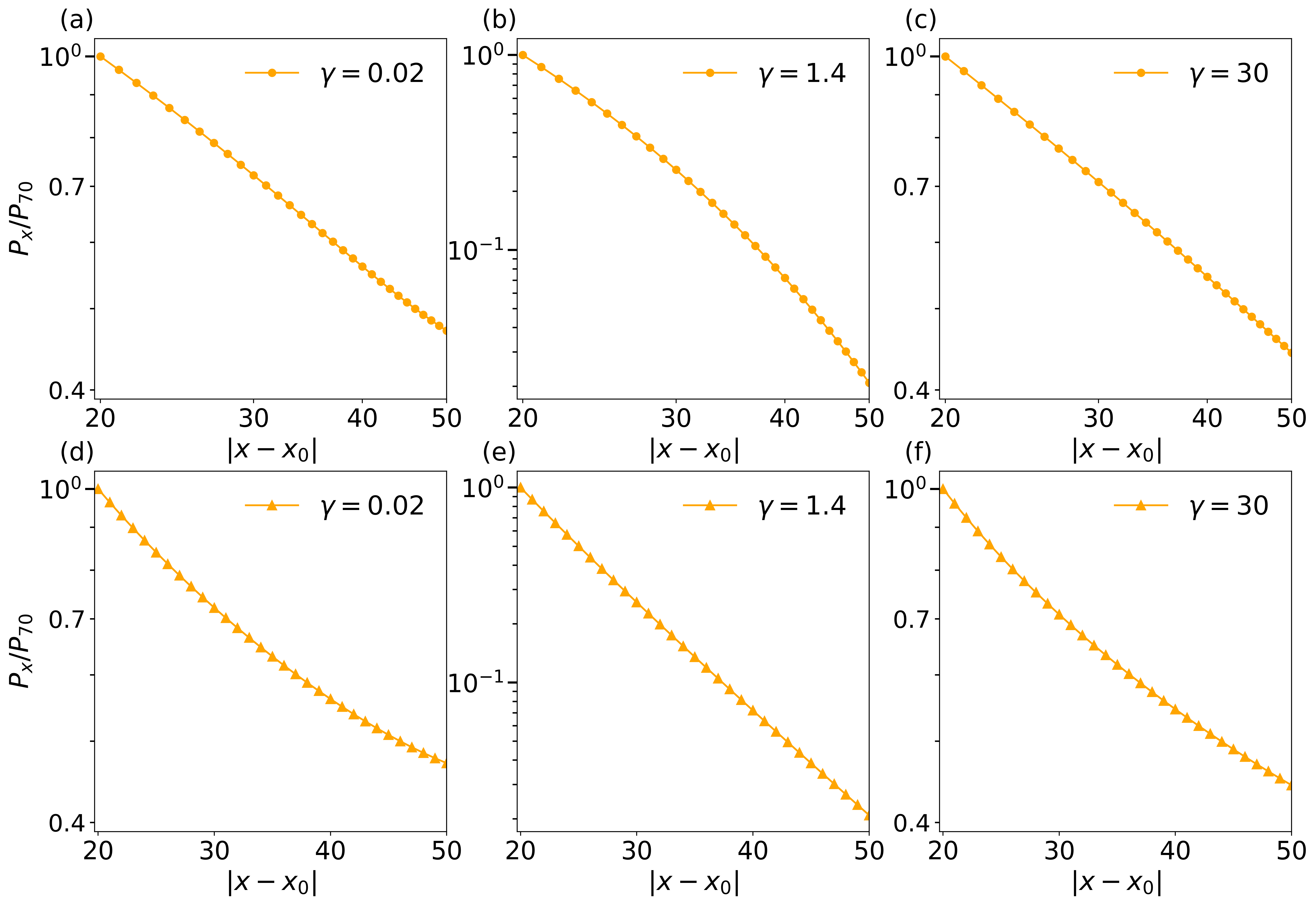}
    \caption{Scale of the bulk loss probability with respect to relative distance at (a) and (d) $\gamma=0.02$, (b) and (e) $\gamma=1.4$, and (c) and (f) $\gamma=30$. A double-logarithmic coordinate is used in (a)-(c), and  a single-logarithmic coordinate is used in (d)-(f). Other parameters are $t_1=0.7,t_2=0.5,L=100$, and $x_0=90.$} 
    \label{fig-5}
\end{figure}

Beyond the consistent region, an intriguing observation is that the pronounced edge loss probability in Fig.~\ref{fig-4}(a) emerges at both large and small values of 
$\gamma$. As illustrated in Fig.~\ref{fig-4}(d), when 
$\gamma=10$
, despite the presence of strong boundary scattering, the edge loss probability is significantly larger than 
$P_{\min}$. In fact, extreme dissipation can enhance scattering [see Fig.~\ref{fig-4}(b)] while simultaneously driving $r_G$ and $\text{min}[|\beta_L(\omega)|]$ toward 1, which corresponds to an overlap between the PBC spectrum and the OBC spectrum. In other words, some of the PBC eigenmodes approach the real axis with little imaginary value, similar to the eigenmodes around  $\omega_0$ discussed in section~\ref{se:IGC}. In summary, extreme dissipation induces a transition from the condition of imaginary gap opening to imaginary gap closing. This transition does not induce edge bursts in the infinite limit, but it becomes feasible in finite-size chains. 
The effects of this transition are more clearly revealed through the behavior of the bulk propagation mode, as shown in Fig.~\ref{fig-5}. 
For moderate values of $\gamma=1.4$, the bulk loss probability  relative to $P_{70}$ follows  $\ln{P_x} \propto \left|x-x_0 \right|$, manifesting as a linear dependence in single-logarithmic coordinates, 
as shown in Fig.~\ref{fig-5}(e). However, under extreme dissipation $\gamma=0.02$ and $\gamma=30$, a linear dependence of the bulk loss probability, $\ln P_x \propto \ln\left|x-x_0 \right|$, is found in  double-logarithmic coordinate. These results return to Eq.~\eqref{eq-bulk}, thereby enabling the emergence of edge bursts. 
In summary, the edge burst here originates from the competition between bulk propagation and boundary scattering: The dominance of the former in certain parameter regions accounts for the reemergence of edge burst.

\section{Edge loss probability in different system sizes}

\begin{figure}
    \centering    \includegraphics[width=1\linewidth]{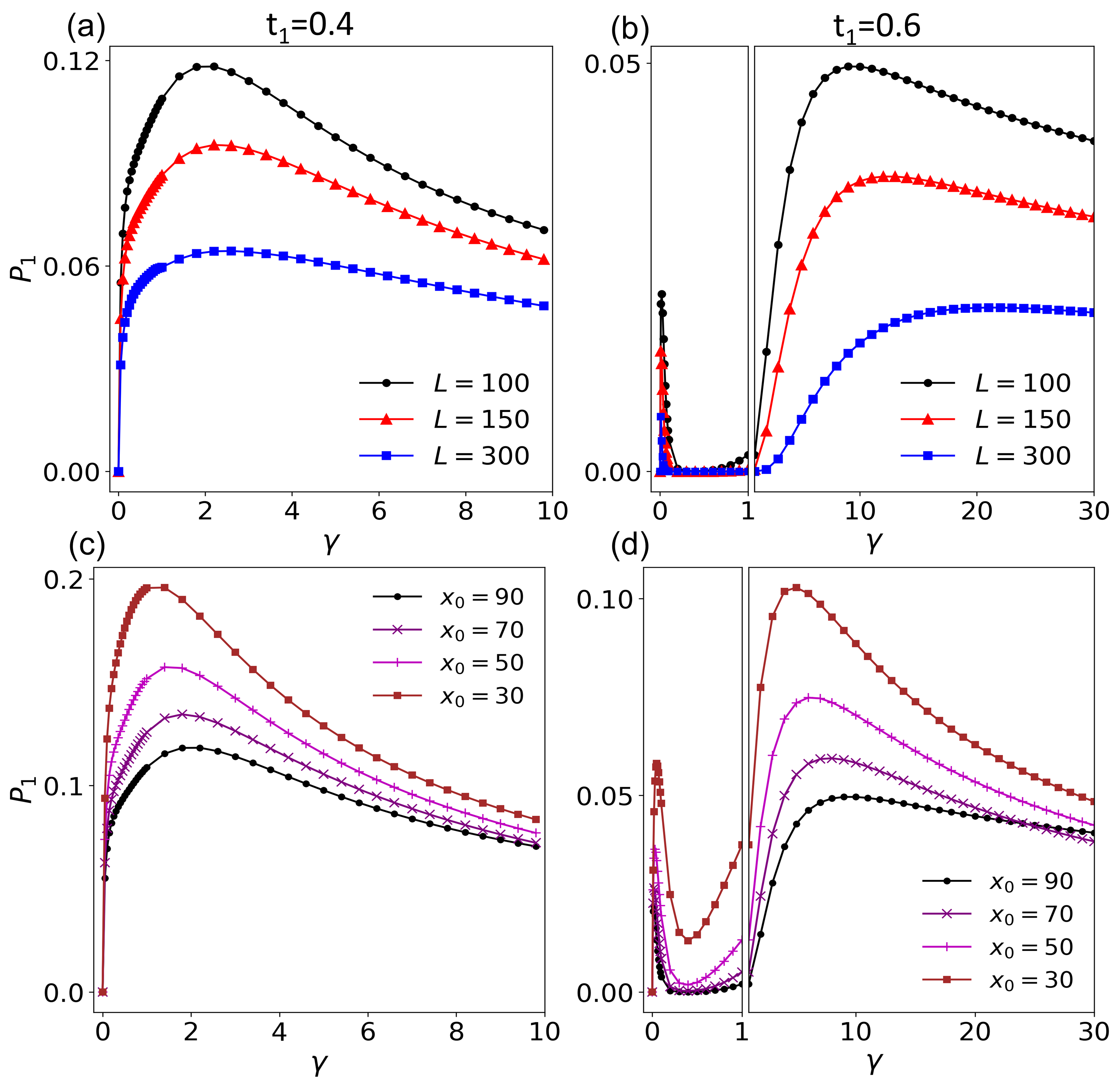}
    \caption{The variation of the edge loss probability with on-site dissipation for (a) and (b) different system size $L=100, 150, 300$ and $x_0=0.9L$ and (c) and (d) different excitation position $x_0=30, 50, 70, 90$ and $L=100$. In (a) and (c), the imaginary gap is closed, $t_1=0.4$ and $t_2 =0.5$, and in and (b) and (d) it is open, $t_1=0.6$ and $t_2 =0.5$.   }
    \label{fig6}
\end{figure}

To further explore the finite-size effects, we systematically investigate the edge loss probability for different system sizes and excitation positions, as shown in Fig.~\ref{fig6}. The corresponding code and data are available online~\cite{code2026}. We considered two representative values of $t_1$: $t_1=0.4$ (where the imaginary gap is closed) and $t_1=0.6$ (where the imaginary gap is open). As illustrated in Fig.~\ref{fig6}, regardless of whether the imaginary gap is closed or open, varying the system size and excitation position does not alter the overall trend of the edge loss probability, 
although the peak position exhibits a slight finite-size dependence. A more notable difference is the magnitude of $P_1$. When the excitation point is placed closer to the edge, the edge loss probability becomes larger because the wave packet experiences less bulk dissipation during propagation. These consistent trends demonstrate the robustness of our key findings—namely, boundary scattering can suppress the edge burst, while strong dissipation combined with NHSE can induce a pronounced edge loss even when the imaginary gap is open. It should also be noted that, except in regimes where boundary confinement is extremely strong, boundary scattering should be considered in experiment and numerical calculations. 

\begin{figure}[htbp]
    \centering    \includegraphics[width=1\linewidth]{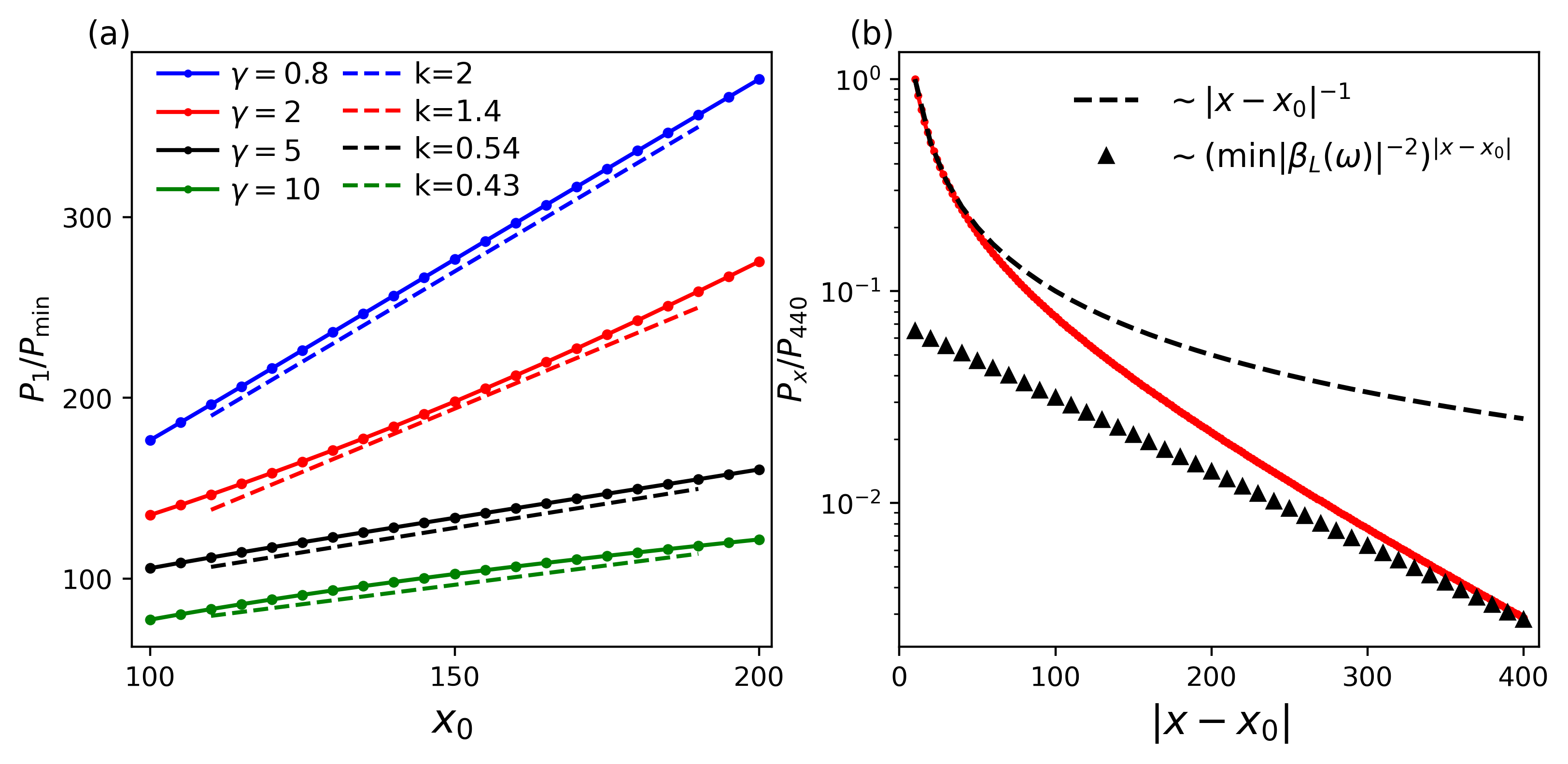}
    \caption{(a) Relative height $P_1/P_{\min}$ as a function of $x_0$ under the condition of imaginary gap closing, $t_1 = 0.4$ and $t_2 = 0.5$, with $L = 250$ for different values of $\gamma$. Here, the numerical results and fitted curves are represented  by lines with dots and dashed lines,
    respectively. (b) The bulk probability distribution of $P_x$ normalized by $P_{440}$  under the condition of imaginary gap opening, $t_1=0.6$ and $t_2=0.5$, with $\gamma=5,x_0=450, L=500$. }
    \label{fig7}
\end{figure}

Although the edge loss probability for the finite-size lattice, especially under strong dissipation, appears to exhibit behavior inconsistent with the non-Hermitian edge burst theory developed in the thermodynamic limit, the finite-size results can be extrapolated to the infinite-length case, as shown in Fig.~\ref{fig7}.  Under the condition of imaginary gap closing with a negligible boundary effect, the ratio
\begin{align}
    P_1/P_{\min}\approx \frac{\frac{4\gamma}{\pi}\Gamma (\frac{3}{2})Q(2K)^{-\frac{3}{2}}x_{0}^{-\frac{1}{2}}}{\frac{2\gamma}{\pi}\Gamma (\frac{3}{2})Q(2K)^{-\frac{3}{2}}|x-x_0|^{-\frac{3}{2}}}\approx 2x_0,
\end{align}
as confirmed by the case with $\gamma=0.8$ in Fig.~\ref{fig7}(a). As $\gamma$ gradually increases, boundary scattering becomes stronger, leading to a small $P_1/P_{\min}$ in finite-size systems and a lower slope of this ratio with respect to $x_0$. However, as $x_0$ (which
can also be considered the system's size) increases, $P_1/P_{\min}$ still increases linearly with the distance between the boundary and the excitation point.  It becomes appreciable at sufficiently large sizes, thereby recovering the edge burst behavior in the infinite-length limit.

Under the condition of imaginary gap opening with strong dissipation, we focus on another signature closely connected to the edge burst: the scaling relation of the bulk loss probability. As shown in Fig.~\ref{fig7}(b), we consider the bulk loss probability within a long chain. Here, the bulk near the excitation point represents the finite-size bulk, while the part far from the excitation point exhibits the bulk behavior in a long system. It can be seen that near the excitation point, $P_x$ normalized by $P_{440}$ exhibits the scaling relation $\ln{P_x} \propto \ln|x-x_0 |$ because $\min |\beta_L(\omega)|\approx 1$ under extreme dissipation, like in Fig.~\ref{fig-5}. However, far from the excitation point, Eq.~(\ref{eq-P1}) yields an exponential law $P_x\sim (\min |\beta_L(\omega)|^{-2})^{|x-x_0|}$, with $\ln{P_x} \propto \left|x-x_0 \right|$; hence, one can reasonably expect the edge
burst to vanish in the infinite-length limit under the condition
of imaginary gap opening.

\section{Discussion and Conclusions}
In conclusion, we investigate the boundary effects on the edge loss probability a in finite-size non-Hermitian lossy lattice. On one hand, in the presence of a strong non-Hermitian skin effect accompanied by a large derivative of the Lyapunov exponent at zero, the boundary exhibits weak scattering but strong confinement. In this regime, the power-law bulk modes, arising from either the condition of imaginary gap
closing or imaginary gap opening under extreme dissipation, are confined by the boundary, leading to edge burst. On the other hand, when the derivative of the Lyapunov exponent at zero is small, the boundary induces significant reflection. As a result, even if the conditions for an edge burst are satisfied in the infinite limit, boundary scattering in a finite-size system can suppress the edge loss probability. 

Our work complements the investigation of the edge burst
by addressing it from the boundary perspective and provides essential insights for experimental realizations, as experimental systems are inherently finite-sized. Consequently, boundary scattering can be clearly observed in all experimental setups (e.g., optical lattices, acoustic lattices) and numerical simulations whenever the boundary confinement is not overwhelmingly strong. Finally, exploring either quantum walks under generalized boundary conditions~\cite{PhysRevLett.127.116801} that interpolate between PBCs and OBCs or non-Hermitian systems with size‑dependent coupling or scale‑free critical behavior~\cite{li2020critical} may reveal richer phenomena concerning edge loss probability and boundary effects.\\

\noindent \textbf{Conflict of interest} 

\vspace{10pt}

The authors declare that they have no conflict of interest.

\vspace{10pt}
\noindent \textbf{Acknowledgments} 

\vspace{10pt}

This work is supported by the
Natural Science Foundation of Hunan Province (Grant No. 2024JJ6011) and the Research Foundation of Education Bureau of Hunan Province (Grant No. 23B0622).
\end{CJK}
\bibliography{manuscript.bib}
\end{document}